# A Survey of the Potential Long-term Impacts of AI

How AI Could Lead to Long-term Changes in Science, Cooperation, Power, Epistemics and Values

Sam Clarke
Leverhulme Centre for the Future of Intelligence
University of Cambridge
Cambridge, United Kingdom
samckclarke@gmail.com

Jess Whittlestone
Centre for the Study of Existential Risk
University of Cambridge
Cambridge, United Kingdom
jlw84@cam.ac.uk

## ABSTRACT

It is increasingly recognised that advances in artificial intelligence could have large and long-lasting impacts on society. However, what form those impacts will take, just how large and long-lasting they will be, and whether they will ultimately be positive or negative for humanity, is far from clear. Based on surveying literature on the societal impacts of AI, we identify and discuss five potential long-term impacts of AI: how AI could lead to long-term changes in science, cooperation, power, epistemics, and values. We review the state of existing research in each of these areas and highlight priority questions for future research.

## 1 Introduction

Artificial intelligence (AI) is already being applied in and impacting many important sectors in society, including healthcare [37], finance [19] and law enforcement [58]. Some of these impacts have been positive—such as the ability to predict the risk of breast cancer from mammograms more accurately than human radiologists [48]—whilst others have been extremely harmful—such as the use of facial recognition technology to surveil Uighur and other minority populations in China [66]. As investment into AI research continues, we are likely to see substantial progress in AI capabilities and their potential applications, precipitating even greater societal impacts.

What is unclear is just how large and long-lasting the impacts of AI will be, and whether they will ultimately be positive or negative for humanity. In this paper we are particularly concerned with impacts of AI on the far future:[1] impacts that would be felt not only by our generation or the next, but by many future generations who could come after us. We will refer to such impacts as *long-term impacts*.

Broadly speaking, we might expect AI to have long-term impacts because of its potential as a *general purpose technology*: one which will probably see unusually widespread use, tend to spawn complementary innovations, and have a large inherent potential for technical improvement. Historically, general purposes technologies—such as the steam engine and electricity—have tended to precipitate outsized societal impacts [28]. In this paper, we consider potential long impacts of AI which could:

- Make a **global catastrophe** more or less likely (i.e. a catastrophe that poses serious damage to human well-being on a global scale, for example by enabling the discovery of a pathogen that kills hundreds of millions of people).[2]
- Make premature **human extinction** more or less likely.
- Make it more or less likely that important aspects of the world are "**locked in**" for an extremely long time (for

---

[1] Following Greaves & MacAskill [32], we take "the far future" to mean everything from some time $t$ onwards, where $t$ is a surprisingly long time from the point of decision (to develop or deploy some AI technology)—say, 100 years.
[2] See Beckstead [4] for arguments that global catastrophes could impact the far future.

example by enabling a robust totalitarian regime to persist for an extremely long time).[3]
- Lead to **other kinds of "trajectory change"** (i.e. other persistent changes which affect how good the world is at every point in the far future, such as eliminating all disease).[4]

We surveyed papers on the societal impacts of AI broadly, identified potential impacts in the four above categories, clustered them into areas, and did further research on each area. In what follows, we discuss five such areas: scientific progress, conflict and cooperation, power and inequality, epistemic processes and problem solving, and the values which steer humanity's future. In each, we will review existing arguments that AI could have long-term impacts, discuss ways in which these impacts could be positive or negative, and highlight priority questions for future research. We conclude with some overall reflections on AI's potential long-term impacts in these areas.

## 2 Scientific progress

One way AI could have long-term impacts is by changing how scientific progress occurs. Most of the most dramatic changes in humanity's trajectory so far can be attributed at least partly to scientific or technological breakthroughs, for example the development of the steam engine enabling industrialisation; many advances in modern medicine leading to much longer lifespans; and the invention of computers and the internet which have fundamentally changed the way we communicate and connect across the world. Of all the ways AI could shape scientific progress, the potential to drastically *speed up* the rate at which progress and breakthroughs occur is perhaps the most important.

### 2.1 How AI could accelerate scientific progress

There are a few different ways AI could enable faster scientific progress:

- **AI systems could help scientists to become more productive**. For instance, consider Ought's Elicit [9] an automated research assistant, powered by the large language model GPT-3 [7]. It has several tools to help researchers do their work faster, e.g. tools for helping with literature reviews, brainstorming research questions, and so on.
- **AI could increase the number of (human) scientists**. As AI systems substitute for human labour in tasks involved in the production of consumer goods, this frees up labour to do other things, including science.[5]

---

[3] We borrow this concept from Greaves & MacAskill [32].
[4] Some have argued that another category of impact on the far future is the speeding up of various kinds of development (e.g., economic, scientific, legal, moral), but we find these arguments less certain so we will not discuss such impacts here.
[5] Here, and in the next point, AI systems need not be directly involved in science themselves.

- **AI could increase the amount of science funding**. By sustaining or increasing economic growth, AI progress could result in more capital available to spend on science.
- **AI could make it possible to automate an increased proportion of the scientific process**. That is, we may manage to train AI systems which can (perhaps fully) automate tasks involved in the scientific process, such as generating new ideas and running experiments.

Of these, the fourth—AI systems automating the scientific process—seems like it could have particularly dramatic impacts. If significant parts of the scientific process could be automated by software that is easily replicated, this could lead to very rapid progress. For example, in fields where running large numbers of experiments or testing many hypotheses is necessary, being able to run tests on software that can be parallelised and run at all hours could speed progress up enormously compared to relying on human scientists.

In some cases, this might look more like speeding up breakthroughs that would eventually have been made by human scientists, but in others, it might make it possible to overcome bottlenecks that were previously intractable (at least for all practical purposes). For example, biologists have been struggling for years to make progress on the protein folding problem, because the space of hypotheses was too large to be tractable for human scientists. The AI system AlphaFold 2 made it possible to test a very large number of hypotheses (each hypothesis being a set of neural network parameters), and led to breakthrough progress on the protein folding problem [38]. At the extreme, complete automation of the scientific process could lead to extremely rapid scientific progress [40].

How long-term would the impacts of faster progress be? If AI enables scientific breakthroughs that *wouldn't otherwise have been possible*—such as eliminating all disease, perhaps—that could constitute a trajectory change. In some cases, merely allowing breakthroughs to occur sooner than they would have otherwise could have very long-lasting impacts, particularly if those breakthroughs make it possible to avert a large threat such as a global pandemic. Beyond this, it seems that what, and how many, kinds of breakthroughs would have long-term impacts is an open question.

Of course, saying that AI is likely to speed up scientific progress, and that such progress could have long-term impacts on society, says nothing about whether those impacts will be overall positive or negative.

### 2.2 Potential benefits

Scientific progress is at least partly to thank for many of the successes of human history so far. Medicine provides the most obvious examples: vaccines, antibiotics and anaesthetic. The discovery of electricity raised living standards in many parts of the world, thanks to products it enabled like washing machines, lightbulbs, and telephones [31]. Moreover, we are already seeing some positive scientific contributions enabled by AI: AlphaFold 2 promises to help with developing treatments for diseases or finding enzymes that break down industrial waste [38]. The use of AI to advance drug discovery is receiving increasing attention [55]: the first clinical trial of an AI-designed drug began in Japan [8] and a number of startups in this space raised substantial funds in 2020 [67].

More importantly for our purposes, there are some potential scientific breakthroughs that could have long-term impacts. For instance, novel energy generation or storage technology could reduce the probability of globally catastrophic climate change. Understanding the ageing process could help slow or reverse ageing [80], which could make the world better at every point in the future (a trajectory change).

## 2.3 Potential risks

On the other hand, faster science and technology progress could make it easier to develop technologies that make a global catastrophe more likely. For example, AI could speed up progress in biotechnology [52, 73], making it easier to engineer or synthesise dangerous pathogens with relatively little expertise and readily available materials. More speculatively, AI might enable progress towards atomically precise manufacturing (APM) technologies,[6] which could make it substantially easier to develop dangerous weapons at scale,[7] or even be misused to create tiny self-replicating machines which outcompete organic life and rapidly consume earth's resources [59].

Exacerbating these problems is that faster scientific progress would make it even harder for governance to keep pace with the deployment of new technologies. When these technologies are especially powerful or dangerous, such as those discussed above, insufficient governance can magnify their harms.[8] This is known as the *pacing problem*, and it is an issue that technology governance already faces [47], for a variety of reasons:

- Information asymmetries between the developers of new technologies and those governing them, leading to insufficient or misguided governance.
- Tech companies are often just better resourced than governments, especially because they can afford to pay much higher salaries and so attract top talent.
- Technology interest groups often lobby to preserve aspects of the status quo that are benefiting them (e.g., subsidies, tax loopholes, protective trade measures), making policy change—and especially experimental policies—difficult and slow to implement [56].
- For governance to keep pace with technological progress, this tends to require anticipating the impacts of technology in advance, before shaping them becomes expensive, difficult and time consuming, and/or catastrophic harms have already occurred. But anticipating impacts in advance is hard, especially for transformative technologies. This is commonly referred to as the Collingridge Dilemma [13].

## 2.4 Open questions

A key question here is: what type and level of AI capabilities would be needed to automate a significant part of the scientific process in different domains? Some areas of scientific research may be easier to automate with plausible advances in AI capabilities, and some types of scientific breakthroughs will be more impactful—whether positive or negative—than others. If we had a better understanding of the types of AI capabilities needed to automate progress in different areas of science, then we could ask about the impacts of progress in areas where automation seems like it might not be too far off.

The possibility of AI speeding up scientific progress also requires us to confront complex questions about what kinds of progress are good for society. For example, AI-enabled progress in cognitive science could make it possible to create digital people: computer simulations of people living in a virtual environment, who could plausibly be as conscious as we are and can do most of the things that humans can. This would totally transform the world as we know it. Whether a world with digital people in it would be better or worse than today's world is an open question depending on many normative and empirical assumptions, worth more serious consideration.

## 3 Cooperation and Conflict

Another way AI could cause long-term impacts is by changing the nature or likelihood of cooperation and conflict between powerful actors in the world, including by:

- Enabling the development of new tools or technologies relevant to cooperation and conflict, such as new tools for negotiation, or new weapons.
- Enabling the automation of decision-making in conflict scenarios, leading to unintentional escalation or otherwise making mistakes or high-risk decisions more likely.
- Altering the strategic decision landscape faced by powerful actors.

## 3.1 How AI could improve cooperation

The potential uses of AI to improve cooperation have not been explored in depth, but a few ideas are covered in Dafoe et al. [17]:

- AI research in areas such as machine translation could enable richer communication across countries and cultures, therefore facilitating the finding of common ground.
- AI methods could be used to build mechanisms to incentivise truthful information sharing.
- AI could help develop languages for specifying commitment contracts (for instance, imagine the potential of assurance

---

[6] APM is a proposed technology for assembling macroscopic objects defined by data files by using very small parts to build the objects with atomic precision using earth-abundant materials [59]. It has not yet been developed, and its feasibility is unclear.
[7] APM could lead to the development of new kinds of drones and centrifuges for enriching uranium that are cheaper and easier to produce [59].
[8] For a historical analogy, consider the cluster of deaths that resulted from the unregulated proliferation of A/C power lines in New York City in the late 19th century [68].

contracts for nuclear disarmament), and improve our ability to reason about the strategic impacts of commitment.
- AI research could explore the space of distributed institutions that promote desirable global behaviours and design algorithms that can predict which norms will have the best properties.

There are a few reasons to expect more cooperation to improve long-run outcomes. The availability of cooperative solutions tends to reduce the likelihood of conflict (of which we will discuss the long-term importance in the next section), costly as it is for all parties involved. Much greater global cooperation than exists today is likely to be crucial for ensuring humanity has a flourishing long-term future,[9] as well as improving collective problem solving more broadly (we'll discuss the latter in more detail in a later section).

### 3.2 How AI could worsen conflict

Equally, AI could have significant impacts on the likelihood and nature of conflict, and in particular could make globally catastrophic outcomes from conflict more likely.

There are several reasons to think this. AI is already enabling the development of weapons which could cause mass destruction—including new weapons that themselves use AI capabilities, such as Lethal Autonomous Weapons [2],[10] and the potential use of AI to speed up the development of other potentially dangerous technologies, such as engineered pathogens (as discussed in Section 2).

Automation of military decision-making could introduce new and more catastrophic sources of error (especially if there are competitive pressures which lead to premature automation). One concern here is humans not remaining in the loop for some military decisions, creating the possibility of unintentional escalation because of:

- Automated tactical decision-making, by 'in-theatre' AI systems (e.g. border patrol systems start accidentally firing on one another), leading to either: tactical-level war crimes,[11] or strategic-level decisions to initiate conflict or escalate to a higher level of intensity—for example, countervalue (e.g. city-) targeting, or going nuclear [62].
- Automated strategic decision-making, by 'out-of-theatre' AI systems—for example, conflict prediction or strategic planning systems giving a faulty 'imminent attack' warning [20].

Furthermore, even if humans remain in the loop, automation is likely to increase the pace and complexity of military decision-making, which could make mistakes or high-risk decisions more likely.

A further concern is that AI could more broadly influence the strategic decision landscape faced by actors in a way that makes conflict more likely or undermines stability. For example, AI could undermine nuclear strategic stability by making it easier to discover and destroy previously secure nuclear launch facilities [30, 46, 49]. AI may also offer more extreme first-strike advantages or novel destructive capabilities that could disrupt deterrence, such as cyber capabilities being used to knock out opponents' nuclear command and control [15, 29]. The use of AI capabilities may make it less clear where attacks originate from, making it easier for aggressors to obfuscate an attack, and therefore reducing the costs of initiating one. By making it more difficult to explain their military decisions, AI may give states a carte blanche to act more aggressively [20]. By creating a wider and more vulnerable attack surface, AI-related infrastructure may make war more tempting by lowering the cost of offensive action (for example, it might be sufficient to attack just data centres to do substantial harm), or by creating a 'use-them-or-lose-them' dynamic around powerful yet vulnerable military AI systems. In this way, AI could exacerbate the 'capability-vulnerability paradox' [22], where the very digital technologies that make militaries effective on the battlefield also introduce critical new vulnerabilities. AI development may itself become a new flash point for conflicts—causing more conflict to occur—especially conflicts over AI-relevant resources (such as data centres, semiconductor manufacturing facilities and raw materials).

Alongside the possibility that AI will make globally catastrophic outcomes from conflict more likely, conflict is in general a destabilising factor which reduces our ability to mitigate other potential global catastrophes and steer towards a flourishing future for humanity. For instance, conflict tends to erode international trust and cooperation, and increases risks posed by a range of weapon technologies [54].

### 3.3 Open questions

It would be good to have more analysis of the kind of AI systems we could develop to help with cooperation. How might we make the development and deployment of such systems more likely?

Likewise, a more detailed understanding of the kinds of military decisions that are likely to be automated, mistakes that might arise, and incentives that will develop, seems very valuable. In what scenarios might AI-enabled warfare lead to unintentional escalation, and how might we prepare to avoid this happening in advance? Are there types or uses of AI systems that we might want to prohibit or seriously restrict because the risks they pose to conflict and international stability are too great?

---

[9] We'll argue this point in Section 6. Basically, the idea is that steering towards a flourishing future requires a high degree of cooperation and coordination, because without it, individual actors must (on pain of being outcompeted) spend a fraction of their resources investing in military and economic competitiveness, rather than creating a flourishing world according to their values.
[10] Whilst we find it hard to tell a plausible story where LAWs lead to a global catastrophe directly, the proliferation of LAWs seems like it would heighten the risk of a global catastrophe, and successful LAWs governance would be a valuable dry-run and precedent for governance of advanced AI.
[11] However, we don't think these could constitute a global catastrophe.

# 4 Power and inequality

It seems likely that AI development will shift the balance of power in the world: as AI systems become more and more capable, they will give those with access to them greater influence, and as AI becomes more integrated into the economy, it will change how wealth is created and distributed. What is not clear is whether the trend will be towards a more or less equal society, and how drastic and long-lasting these power shifts might be. For the purposes of this section, we will be talking specifically about (in)equality in political power and wealth.

## 4.1 How AI could reduce inequality

It is plausible that AI will increase economic growth rates, and advanced AI could significantly increase them [72]. Whilst the connection between economic growth and inequality isn't clear, there is evidence that poverty has reduced with economic growth in developing countries [23].[12] Thus, it's plausible that the wealth and abundance of resources generated by AI will precipitate significant poverty reduction.

Furthermore, AI could help with identifying and mitigating sources of inequality directly [75]. Some early-stage suggestions include simulating how societies may respond to changes [65], or preventing discrimination in the targeting of job advertisements [18]. That said, we found it particularly difficult to find literature on how AI could reduce inequality in a lasting way, and we think this area could do with more attention.

## 4.2 How AI could exacerbate inequality

There are several ways we might be concerned about the development and use of AI increasing power concentration or inequality. We're already seeing some very concerning trends:

- AI-driven industries seem likely to tend towards monopoly and could result in huge economic gains for a few actors: there seems to be a feedback loop whereby actors with access to more AI-relevant resources (e.g., data, computing power, talent) are able to build more effective digital products and services, claim a greater market share, and therefore be well-positioned to amass more of the relevant resources [14, 39, 45]. Similarly, wealthier countries able to invest more in AI development are likely to reap economic benefits more quickly than developing economies, potentially widening the gap between them.
- The harms and benefits of AI are likely to be very unequally distributed across society: AI systems are already having discriminatory impacts on marginalised groups [1, 78] and these groups are also less likely to be in a position to benefit from advances in AI such as personalised healthcare [78].
- AI-based automation has the potential to drastically increase income inequality. It seems quite plausible that progress in reinforcement learning and language models specifically could make it possible to automate a large amount of manual labour and knowledge work respectively [35, 45, 69], leading to widespread unemployment, and the wages for many remaining jobs being driven down by increased supply.
- Developments in AI are giving companies and governments more control over individuals' lives than ever before, and may possibly be used to undermine democratic processes. We are already seeing how the collection of large amounts of personal data can be used to surveil and influence populations, for example the use of facial recognition technology to surveil Uighur and other minority populations in China [66]. Further advances in language modelling could also be used to develop tools that can effectively persuade people of certain claims [42].

Many of the trends described above could combine to create a world which is much more unequal than the one we live in today, both in terms of wealth and political power. If AI is embedded in society in ways that create self-reinforcing feedback loops, whereby those who are already rich and powerful are able to continue reaping benefits of AI, and those who are poor and powerless lack access to the same benefits and are at greater risk of harms, this could make it even more difficult to break cycles of inequality than it is today—making it more likely that inequality will persist for a very long time.

A particularly extreme scenario would be one where AI development enables a relatively small group of people to obtain unprecedented levels of power, and to use this to control and subjugate the rest of the world for a long period of time.

Gaining power in this way could be sudden or gradual. A sudden gain in control could look like some group developing and controlling much more powerful AI systems than anyone else, and using them to gain a decisive strategic advantage: "a level of technological and other advantages sufficient to enable … complete world domination" [5]. A historical analogy for a sudden takeover would be takeovers by Spanish conquistadors in America, at least in part due to their technological and strategic advantages [41].

A gradual gain in control could look like the values of the most advanced actors in AI slowly coming to have a large influence over the rest of the world. For instance, if labour becomes increasingly automated, we could end up in a world where it's very difficult or even impossible to trade labour for income (i.e. labour share of GDP falls to near zero compared to capital share), meaning the future would be controlled by the relatively small proportion of people who own the majority of capital/AI systems [72]. A historical analogy for a gradual takeover would be how WEIRD (Western, Educated, Industrialized, Rich and Democratic) values (e.g. analytical, individualistic thinking; less clannishness and more trusting in abstract rules and intermediating institutions) gradually came to dominate large parts of the world because they were more (economically and militarily) "successful" [34].

In addition to making it easier for some groups to *obtain* large amounts of power, developments in AI could also make it easier to *retain* that power over long periods of time. This might either be done very directly—a group might use AI-based surveillance and manipulation to identify and suppress opposition and perpetuate a global totalitarian regime, for instance—or more indirectly—such as a globally powerful group which embeds its values and objectives in powerful AI systems that themselves come to control society (and note that, unlike humans, AI systems can be programmed to reliably pursue the same goals/plans over a long period of time).

---
[12] Adams [2003] measures economic growth via mean income and GDP per capita, and finds statical links between poverty reduction and both of these measures.

### 4.3 Open questions

It would be valuable to survey—and evaluate the effectiveness and feasibility of—governance tools for reducing the likelihood and severity of AI-induced power concentration and inequality. Some suggestions here include the "windfall clause" [53], and distributing ownership of companies (especially AI-assisted ones) and land [50]. After surveying existing tools, one could think about potential modifications to make them more effective and feasible, or try to identify new tools or strategies. Previous waves of automation have had differing effects on inequality [26], so it could also be valuable to look for generalisable lessons: what regulation, institutions or other governance structures have been successful at promoting more equal outcomes in the face of technological change?

## 5 Epistemic processes and problem solving

Our ability to solve problems and make progress as a society towards a great long-term future depends heavily on epistemic processes: how information is produced and distributed, and the tools and processes we use to make decisions and evaluate claims.

AI development is already impacting both of these things, and may therefore significantly shape our ability to solve problems as a society far into the future—for better or worse. Currently, almost all information on the internet is created by humans, but we are beginning to see this change, as AI-generated content becomes more convincing [74]. As it becomes a significant fraction of the information available, its purpose and quality could have a significant impact on what we believe and how we solve problems, individually and collectively. AI systems and tools are at the same time playing an increasingly large role in how we filter, process, and evaluate information, again shaping the way individuals and communities view the world [77].

### 5.1 How AI could improve epistemic processes

AI could help us understand complex aspects of the world in a way that makes it easier to identify and mitigate threats to humanity's long-term future. For instance, AI is already used to support early warning systems for disease outbreaks: machine learning algorithms were used to characterise and predict the transmission patterns of both Zika virus [36] and SARS-CoV-2 [21, 79], supporting more timely planning and policymaking. With better data and more sophisticated systems in the future it may be possible to identify and mitigate such outbreaks much earlier [63].

If developments in AI could be leveraged to enable better cooperation between groups, as mentioned earlier, this could also make it much easier to solve global problems. For instance, AI could help different groups to make verifiable claims that they are minimising negative externalities from activities like biotechnology research (e.g. via AI-enabled surveillance of the riskiest biotechnology labs).

Along with mitigating threats and enabling cooperation to solve global problems, AI-based tools to support better reasoning could help humans far surpass the amount of intellectual progress we could otherwise have made on important problems. It seems that steering towards a great future requires making progress on difficult intellectual problems, like ethics, AI alignment and group decision making, and that AI-based tools to facilitate reasoning about these problems could be of central importance. Ought's Elicit (the AI-based research assistant mentioned in Section 2.1) is a suggestive example.

### 5.2 How AI could worsen epistemic processes

However, as AI plays an increasingly large role in how information is produced and disseminated in society, this could also distort epistemic processes and undermine collective solving capacities. One of the most significant commercial uses of current AI systems is in the content recommendation algorithms of social media companies, and there are already concerns that this is contributing to worsened polarisation online [24, 57]. At the same time, we are seeing how AI can be used to scale up the production of convincing yet false or misleading information online (e.g. via image, audio, and text synthesis models like BigGAN [6] and GPT-3 [7]).

As AI capabilities advance, they may be used to develop sophisticated persuasion tools, such as those that tailor their communication to specific users to persuade them of certain claims [42]. While these tools could be used for social good—such as New York Times' chatbot that helps users to persuade people to get vaccinated against Covid-19 [27]—there are also many ways they could be misused by self-interested groups to gain influence and/or to promote harmful ideologies.

Even without deliberate misuse, widespread use of powerful persuasion tools could have negative impacts. If such tools were used by many different groups to advance many different ideas, we could see the world splintering into isolated "epistemic communities", with little room for dialogue or transfer between communities. A similar scenario could emerge via the increasing personalisation of people's online experiences—in other words, we may see a continuation of the trend towards "filter bubbles" and "echo chambers", driven by content selection algorithms, that some argue is already happening [3, 25, 51].

In addition, the increased awareness of these trends in information production and distribution could make it harder for anyone to evaluate the trustworthiness of any information source, reducing overall trust in information.

In all of these scenarios, it would be much harder for humanity to make good decisions on important issues,

particularly due to declining trust in credible multipartisan sources, which could hamper attempts at cooperation and collective action. The vaccine and mask hesitancy that exacerbated Covid-19, for example, were likely the result of insufficient trust in public health advice [71]. These concerns could be especially worrying if they play out during another major world crisis. We could imagine an even more virulent pandemic, where actors exploit the opportunity to spread misinformation and disinformation to further their own ends. This could lead to dangerous practices, a significantly increased burden on health services, and much more catastrophic outcomes [64].

## 5.3 Open questions

It would be valuable to better understand the kind of AI systems we could develop to improve society's epistemic processes. There are also important questions relating to governance: what governance levers are available for reducing the risk of persuasion tools and online personalisation undermining epistemic processes? Are there datasets we could collect to help with measuring relevant properties of AI systems, like the extent to which they help rather than persuade their users?

# 6 The values that steer humanity's future

One way or another, advanced AI is likely to have a large impact on the values that steer humanity's future.

So far in human history, it seems like the future has largely been determined by competitive pressures, rather than by deliberate attempts by humans to shape the future according to their values. That is, if some technical, institutional, or cultural "innovation" offers a competitive advantage, then its proliferation is highly likely. Some examples include:

- **Firearms** in the Tokugawa period of Japanese history: up until 1853, firearms technology was largely eliminated in Japan, and a samurai-dominated social order persisted for over 200 years. But then, in order to repel the threat of Western colonisation, Japan was forced to readopt firearms, along with other Western customs and institutions—despite this running contrary to the values of the Japanese elite pre-1853 [16].
- **Agriculture:** hunter-gatherer societies which did not adapt to agriculture were gradually killed off by farming societies, who could grow larger due to increased food production.
- **Industrialisation:** states which did not industrialise after the first Industrial Revolution rapidly fell behind in economic production and national competitiveness. Hence there was competitive pressure to industrialise, even if this meant disregarding certain existing societal values.

However, it's plausible that advanced AI will change this. On the one hand, it could be a significant opportunity for (at least some subset of) humanity to increase their ability to deliberately shape the future according to their values—which we will refer to as the opportunity to gain "greater control over the future".

On the other hand, it could cause humans as a species to lose what potential we have for gaining control over the future.

## 6.1 How AI could increase humanity's ability to shape the future

There are several ways in which advanced AI could help humanity to gain greater control over the future. It's important to note that human values having greater control over the future doesn't necessarily mean that the future will be more desirable than one that's driven more by competitive pressures—this depends on what/whose values become influential and how representative they are of broader interests.

First, as mentioned above, advanced AI could improve humanity's ability to cooperate, therefore helping to overcome competitive pressure as a force shaping our future. For example, if we design AI systems that have superhuman cooperative capabilities (e.g. because they are better than humans at making credible commitments), we could move towards a future which avoids common traps of multiagent situations, like destructive conflict or wasting resources on arms races. This could affect how good the world is at every point in the far future, because the resources that would have been spent on conflict and harmful competition could instead be spent however our descendants deem most valuable.

Second, advanced AI could accelerate moral progress, for example by playing a "Socratic" role in helping us to reach better (moral) decisions ourselves (inspired by the role of deliberative exchange in Socratic philosophy as an aid to develop better moral judgements) [44]. Specifically, such systems could help with providing empirical support for different positions, improving conceptual clarity, understanding argumentative logic, and raising awareness of personal limitations. For an early example of this type, Ought's Elicit (the AI-based research assistant mentioned in Section 2.1) has a "bias buster" tool, which attempts to point out cognitive biases that may be at play in a dilemma its user is facing. Whilst it's unclear in general whether the effects of accelerating moral progress wouldn't simply "wash out,"[13] such tools could have long-term impact if they accelerated progress before morally relevant irreversible decisions are made about, for instance, global norms or institutions, or the environments of digital people.

Third, humanity's control over the future is currently threatened by hazards in our environment, and AI could help to mitigate these. In particular, AI could help us to better understand and mitigate potential global catastrophes such as climate change, including by improving resource management, making it easier to rely on an increasing number of variable energy sources, or even by automating the time-consuming processes of discovering new materials that can better store or harness energy [60].

---

[13] For an example of progress that plausibly "washed out", consider that if Edison hadn't invented the light bulb, then soon after someone else probably would have. So, whilst Edison certainly brought forward the date of invention, it's unclear whether he specifically caused any long-lasting impacts via this invention.

## 6.2 How AI could lead to humans losing control of the future

On the other hand, AI could cause humanity to lose our potential for gaining control over the future. The main concern here is that we might develop advanced AI systems whose goals and values are different from those of humans, and are capable enough to take control of the future away from humanity.

The obvious question is: *why* would we develop advanced AI systems that are willing and able to take control of the future? One major concern is that we don't yet have ways of designing AI systems that reliably do what their designers want. Instead, modern AI training[14] works by (roughly speaking) tweaking a system's "parameters" many times, until it scores highly according to some given "training objective", evaluated on some "training data". For instance, the large language model GPT-3 [7] is trained by (roughly speaking) tweaking its parameters until it scores highly at "predicting the next word" on "text scraped from the internet".

However, this approach gives no guarantee that a system will continue to pursue the training objective as intended over the long run. Indeed, notice that there are many objectives a system could learn that will lead it to score highly on the training objective but which do not lead to desirable behaviour over the long run. For instance:

- The system could learn the objective "maximise the contents of the memory cell where the score is stored" which, over the long run, will lead it to fool the humans scoring its behaviour into thinking that it is doing what they intended, and eventually seize control over that memory cell, and eliminate actors who might try to interfere with this. When the intended task requires performing complex actions in the real world, this alternative strategy would probably allow the system to get much higher scores, much more easily, than successfully performing the task as intended.
- Suppose that some system is being trained to further some company's objective. This system could learn the objective "maximise quarterly revenue" which, over the long run, would lead it to (e.g.) collude with auditors valuing the company's output, fool the company's directors, and eventually ensure no actor who might reduce the company's revenue can interfere.

It's also worth noting that, to the extent that these incorrect objectives are easier to represent, learn, or make plans towards than the intended objective—which is likely, because we will be trying to use AI to achieve difficult tasks—then they may be the kind of objectives that AI systems learn by default.[15]

This kind of behaviour is currently not a big issue, because AI systems do not have very much decision-making power over the world. When failures occur, they look like amusing anecdotes rather than world-ending disasters [43].

But as AI systems become more advanced and begin to take over more important decision-making in the world, an AI system pursuing a different objective from what was intended could have much more worrying consequences.

What might these consequences look like in practice? In one scenario, described by Christiano [11], we gradually use AI to automate more and more decision-making across different sectors (e.g., law enforcement, business strategy, legislation), because AI systems become able to make better and faster decisions than humans in those sectors. There will be competitive pressures to automate decisions, because actors who decide not to do so will fall behind on their objectives and be outcompeted. Regulatory capture by powerful technology companies will also contribute to increasing automation—for example, companies might engage in political donations or lobbying to water down regulation intended to slow down automation.

To see how this scenario could turn catastrophic, let's take the example of AI systems automating law enforcement. Suppose these systems that have been successfully trained to minimise reported crime rate.

Initially, law enforcement would probably seem to be improving. Since we're assuming that automated decision-making is better and faster than human-decision making, reported crime will in fact fall. We will be increasingly depending on automated law enforcement—and investing less in training humans to do the relevant jobs—such that any suggestions to reverse the delegation of decision-making power to AI systems would be met with reasonable concern that we just cannot afford to.

However, reported crime rate is not the same as the true prevalence of crime. As AI systems become more sophisticated, they will continue to drive down reported crime by hiding information about law enforcement failures, supressing complaints, and manipulating citizens.[16] As the gap between how things are and how they appear grows, so too will the deceptive abilities of our automated decision-making systems. Eventually, they will be able to manipulate our perception of the world in sophisticated ways (e.g. highly persuasive media or education), and they may explicitly oppose any attempts to shut them down or modify their objectives—because human attempts to take back influence will result in reported crime rising again, which is precisely what they have been trained to prevent. The end state would be one where automated decision-making—by AI systems with objectives that aren't what we intended—has much more influence over the future than human decision-making.

---

[14] Specifically, we are talking in this section about training cutting-edge deep neural networks.

[15] An analogy here is how humans learned simple proxies for "maximise genetic fitness" like "secure food and other resources" and "reproduce" rather than the actual objective that evolution optimised us for, which is much harder to make plans to achieve.

[16] The use of predictive policing algorithms is already facing a smaller scale version of these kind of failures, for the same reason: these algorithms are designed to minimise reported crime, not actual crime [76].

Of course, if we manage to work out how to train law enforcement AI to minimise *actual* crime, then we will be able to avoid these catastrophic failures. However, we don't yet have any methods for training AI systems to reliably pursue such complex objectives, and instead have to resort to proxies like *reported* crime, that will plausibly lead to the kind of scenario described above. This general concern is often known as the "alignment problem" [10, 61]. Note that there have been various other concrete depictions of how failure to solve this problem could play out, and they can look quite different depending on how many powerful AI systems there are and how rapidly their capabilities improve [12].

Exacerbating this problem is that we don't properly understand how modern AI systems work, so we lack methods for checking if their learned objectives are the ones we intended. "Interpretability" is the branch of machine learning that is trying to make progress on this, but it currently is lagging significantly behind the cutting edge of AI capabilities.

A second, less explored concern is that AI could affect competitive pressure as a force shaping the future, in a way that leaves humanity more powerless even if we don't explicitly "lose control" to AI systems. For example, suppose AI accelerates certain kinds of scientific research (especially cognitive science), enabling the creation of digital people. Just as people, and groups of people, are subject to competitive pressures today (e.g., individuals working hard and sacrificing other values like family, love and play, or nations investing in weapons technology and sacrificing spending on education and healthcare), digital people, and groups of them, could be subject to the same kinds of competitive pressure. However, the situation with digital people could be even worse: whilst biological people need at least a minimum threshold of wellbeing to survive and continue competing, the same need not be true of digital people. There could be pressure on digital people to "edit" their own preferences to become ever more productive and competitive. That is, with biological humans, there is a "floor" to the amount of value a future driven by competitive pressure alone could have, but this need not be the case for digital humans. In this scenario, AI does not change the *level* of competitive pressure; it *lowers the floor* on how valueless the future could become. This kind of scenario is discussed by Bostrom [70], mentioned by Dafoe [15] (where it is called "value erosion from competition"), and one version of it is presented in detail by Hanson [33].

### 6.3 Open questions

There are several strategic and governance questions here that it would be very valuable to have a better understanding of. How difficult is the problem of designing advanced AI systems that reliably do what their designers want? Should we expect alignment "warning shots"—i.e. small scale catastrophes resulting from AI systems that do not do what their designers want—which could galvanise more work on the alignment problem? Under the assumption that the alignment problem is very difficult, and there are no warning shots, are there any governance tools that could help avert catastrophe? What governance tools could help avoid "value erosion from competition" scenarios?

## 7 Conclusion

We conclude with some overall reflections on AI's potential long-term impacts in each of the areas we have considered.

**Scientific progress:** AI could lead to very rapid scientific progress which would likely have long-term impacts, but it's very unclear if these would be positive or negative. Much depends on the extent to which risky scientific domains are sped up relative to beneficial or risk-reducing ones, on who uses the technology enabled by this progress, and on how it is governed.

**Cooperation and conflict:** we're seeing more focus and investment on the kinds of AI capabilities that make conflict more likely and severe, rather than those likely to improve cooperation. So, on our current trajectory, AI seems more likely to have negative long-term impacts in this area.

**Power and inequality:** there are a lot of pathways through which AI seems likely to increase power concentration and inequality, though there is little analysis of the potential long-term impacts of these pathways. Nonetheless, AI precipitating more extreme power concentration and inequality than exists today seems a real possibility on current trends.

**Epistemic processes and problem solving:** we currently see more reasons to be concerned about AI worsening society's epistemic processes than reasons to be optimistic about AI helping us better solve problems as a society. For example, increased use of content selection algorithms could drive epistemic insularity and a decline in trust in credible multipartisan sources, which reducing our ability to deal with important long-term threats and challenges such as pandemics and climate change.

**The values that steer humanity's future:** humanity gaining more control over the future due to developments in AI, or losing our potential for gaining control, both seem possible. Much will depend on our ability to solve the alignment problem, who develops powerful AI first, and what they use it for.

These long-term impacts of AI could be hugely important but are currently under-explored. We've attempted to structure some of the discussion and stimulate more research, by reviewing existing arguments and highlighting open questions. While there are many ways AI could in theory enable a flourishing future for humanity, trends of AI development and deployment in practice leave us concerned about long-lasting harms. We would particularly encourage future work that critically explores ways AI could have positive long-term impacts in more depth, such as by enabling greater cooperation or problem-solving around global challenges.

### ACKNOWLEDGMENTS

The authors thank Matthijs Maas, Michael Aird, Shahar Avin, Ben Garfinkel, Richard Ngo, Haydn Belfield, Seán Ó hÉigeartaigh, Charlotte Siegmann, Alexis Carlier and Markus Anderljung for valuable feedback on an earlier version.